\documentclass[a4paper,fleqn]{cas-sc}

\usepackage[authoryear,longnamesfirst]{natbib}

\usepackage{subfigure}
\usepackage{letltxmacro}
\usepackage{hyperref}
\usepackage{amsmath}
\usepackage{booktabs}
\usepackage{longtable}
\usepackage{array}
\usepackage{multirow}
\usepackage{wrapfig}
\usepackage{colortbl}
\usepackage{pdflscape}
\usepackage{tabu}
\usepackage[normalem]{ulem}
\usepackage{makecell}
\usepackage{xcolor}
\usepackage{caption}
\usepackage{tabularx}
\usepackage{adjustbox}
\usepackage{natbib}
\usepackage{mfirstuc}
\usepackage{float}
\usepackage{placeins}
\usepackage{footnote}
\usepackage{stackengine}

\usepackage{ragged2e}

\makeatletter
\newcommand{\customlabel}[2]{%
   \protected@write \@auxout {}{\string \newlabel {#1}{{#2}{\thepage}{#2}{#1}{}} }%
   \hypertarget{#1}{}
}
\makeatother

\bibpunct{(}{)}{,}{a}{}{;}

\DeclareCaptionFormat{bold}{\textbf{#1#2}#3}
\setcitestyle{aysep={}}
\setcitestyle{authoryear,open={(},close={)}}

\def\tsc#1{\csdef{#1}{\textsc{\lowercase{#1}}\xspace}}
\tsc{WGM}
\tsc{QE}
\tsc{EP}
\tsc{PMS}
\tsc{BEC}
\tsc{DE}

\begin{document}
\let\WriteBookmarks\relax
\def\floatpagepagefraction{1}
\def\textpagefraction{.001}

\shorttitle{\textbf{Market Money}}

\shortauthors{Gordon Getty}

\title [mode = title]{\textbf{A separate way to measure rate of return}}                      



%
\author[1]{Gordon Getty}[
                        type=editor,
                        role=Researcher,
                        orcid=0000-0002-0939-6932
                        ]

\cormark[1]
\ead{ggetty@LeakeyFoundation.org}

\author[2]{Nikita Tkachenko}[
                    type=editor,
                    role=Assistant,
                    orcid=0009-0003-8681-3335
]

\ead{natkachenko@usfca.edu}







\affiliation[1]{%
            organization={Leakey Foundation},
            addressline={1003B O’Reilly Avenue}, 
            city={San Francisco},
            postcode={94129}, 
            state={California},
            country={United States}}

\affiliation[2]{organization={Graduate Student, University of San Francisco},
            addressline={2130 Fulton St}, 
            city={San Francisco},
            postcode={94117}, 
            state={California},
            country={United States}}
            





\cortext[cor1]{Corresponding author}



\begin{abstract}
Net profit is sometimes found from data for net operating surplus.  We propose a way to find it from data for consumption, pay and market-value capital, and concomitantly to reveal the factor shares in consumption.
\\
\end{abstract}



\begin{keywords}
Net profit\sep Market-value capital \sep Consumption \sep Pay
\end{keywords}

\maketitle

\section{Introduction}

Many economists are interested in net profit as the capital share of income, and in rate of return as its ratio to capital invested. 
Rate of return to equity and debt investments has been derived from market quotes and dividends, from census methods as to real estate, and from real interest rates.\footnote{See e.g. \cite{jorda2019}, and \cite{homer2005history}.}We suggest that these methods can be supplemented by another which reasons from data for consumption, pay\footnote{Compensation of employees plus a share of mixed income.} and market-value capital as reported in national accounts. We argue, that is,
that cash flow can be found as consumption less pay, and that this difference can then be added to change in market-value capital to give net profit at market value. Net profit can be divided by market-value capital to give rate of return. The argument will meanwhile help clarify factor shares in consumption.

\section{A separate measure of net profit}

National accounts and most teaching are organized on the principles that net output is realized in the forms of consumption and capital growth\footnote{Usually stated as consumption and net investment, but with the understanding that net investment is realized in capital growth. We measure capital at market value, not at its cost in depreciated investment.}, and that this sum equals the factor shares in it as pay and net profit. Thus net output and the equal claims on it show as
\begin{equation}
    C + \Delta K = \Pi + P \quad \text{in national accounts,} \label{eq-1}
\end{equation}
where these notations respectively give consumption, capital growth, pay and net profit. Appendix \ref{appendix-a1} will question the teaching that these two sums give net output and claims on net output. Eq. \eqref{eq-1} in itself, which does not refer to net output or income, will be confirmed nonetheless in Section 5 below by another argument, and in Appendix \ref{appendix-a2} by another still.

Meanwhile Eq. \eqref{eq-1} can be arranged as
\begin{equation}
    C - \Pi = P - \Delta K \ . \label{eq-2}
\end{equation}

$P-\Delta K$, or net profit less capital growth, gives cash flow, or the net flow of value passed from assets to their owners for consumption or reinvestment in other assets. We notate cash flow as $F(K)$, and write\footnote{We find no standard notation for cash flow or cash flow rate, as might be expected in that the cash flow concept is little used in economics outside the specialty of finance. Our choice $F(K)$ risks misinterpretation in that the form $F(X)$ tends to mean a unspecified function of $X$. We use $F(K)$, rather than simply $F$, because we believe that the cash flow from human capital can also be usefully defined, and can show as $F(H)$ (see Appendix \ref{appendix-a1}).}
\begin{equation}
    P - \Delta K = F(K) \ . \quad \text{By Eq. \eqref{eq-2}, then} \label{eq-3}
\end{equation}
\vspace{-5ex}
\begin{equation}
    C - \Pi = F(K) \label{eq-4} \ .
\end{equation}
By Eqs. \eqref{eq-3} and \eqref{eq-4}, then, it is possible to measure net profit as 
\begin{equation}
P = \Delta K + F(K) = \Delta K + C - \Pi\ . \label{eq-5}
\end{equation}
We define
\begin{equation*}
r(K) = \frac{P}{K}\ , \quad g(K) = \frac{\Delta K}{K} \quad \text{and} \quad f(K) = \frac{F(K)}{K} = \frac{C - \Pi}{K} \ ,
\end{equation*}
where $r(K)$ is rate of return, $g(K)$ is capital growth rate, and $f(K)$ is our notation for cash flow rate\footnote{See footnote 4.}. These definitions allow division of Eq. \eqref{eq-5} by capital to show
\begin{equation}
    r(K) = g(K) + f(K) = g(K) + \frac{C-\Pi}{K}\ . \label{eq-6}
\end{equation}
Table \ref{tbl-fgr_table} shows \(f(K), g(K)\) and \(r(K)\) as inferred from data for \(C,\Pi\) and \(K\) through Eqs. \eqref{eq-5} and \eqref{eq-6}.
\FloatBarrier
\begin{table}[pos = H]
    \caption{Average \(f(K)\), \(g(K)\) and \(r(K)\) in all countries, in descending order of net income per capita(\%).}
    \makebox[\textwidth][c]{
    {\centering
        \begin{tabular}{lllll}
 \toprule
Country & Period & $f(K)$ & $g(K)$ & $r(K)$\\
\midrule
Luxembourg & 1996 - 2018 & -0.09 & 4.50 & 4.40\\
Norway & 1981 - 2020 & 2.37 & 5.64 & 8.01\\
Netherlands & 1996 - 2019 & 0.53 & 3.35 & 3.88\\
Switzerland & 1995 - 2019 & 0.60 & 3.70 & 4.30\\
United States & 1970 - 2018 & 3.56 & 3.56 & 7.11\\
\addlinespace
Denmark & 1996 - 2020 & 1.71 & 5.42 & 7.13\\
Austria & 1996 - 2019 & 1.16 & 2.59 & 3.75\\
Sweden & 1950 - 2020 & 1.20 & 4.25 & 5.44\\
Ireland & 1996 - 2019 & 0.90 & 5.61 & 6.52\\
Saudi Arabia & 2002 - 2009 & 11.12 & 5.73 & 16.85\\
\addlinespace
Belgium & 1996 - 2019 & 0.98 & 2.80 & 3.78\\
Iceland & 2000 - 2014 & 2.09 & 3.39 & 5.49\\
Germany & 1990 - 2020 & 2.44 & 3.11 & 5.55\\
Cyprus & 1996 - 2019 & 5.53 & 4.94 & 10.47\\
France & 1950 - 2019 & 1.19 & 4.28 & 5.48\\
\addlinespace
Australia & 1960 - 2019 & 2.03 & 4.45 & 6.49\\
Canada & 1972 - 2020 & 2.20 & 4.45 & 6.65\\
Malta & 1996 - 2019 & 0.99 & 4.01 & 5.00\\
Finland & 1996 - 2020 & 2.45 & 4.06 & 6.51\\
Israel & 2000 - 2019 & 12.16 & 8.75 & 20.91\\
\addlinespace
UK & 1970 - 2019 & 4.03 & 3.00 & 7.04\\
New Zealand & 1999 - 2017 & 3.30 & 5.65 & 8.95\\
Aruba & 1996 - 2000 & -5.61 & 28.80 & 23.19\\
Spain & 1995 - 2019 & 1.97 & 3.57 & 5.53\\
Japan & 1980 - 2017 & 1.74 & 2.81 & 4.56\\
\addlinespace
Slovenia & 1996 - 2019 & -0.17 & 1.72 & 1.54\\
Lithuania & 1996 - 2019 & 6.20 & 4.49 & 10.68\\
Czechia & 1994 - 2019 & 1.03 & 1.07 & 2.10\\
Italy & 1980 - 2020 & 3.99 & 2.84 & 6.83\\
Estonia & 1996 - 2019 & 0.63 & 4.29 & 4.92\\
\addlinespace
Poland & 1996 - 2019 & 4.84 & 5.86 & 10.70\\
Slovakia & 1996 - 2020 & 3.54 & 4.15 & 7.69\\
Hungary & 1996 - 2019 & -0.63 & 3.13 & 2.50\\
Greece & 1996 - 2019 & 22.64 & 0.63 & 23.27\\
\bottomrule \end{tabular}
}

\centering{
\begin{tabular}{lllll}
     \toprule
Country & Period & $f(K)$ & $g(K)$ & $r(K)$\\
\midrule
Portugal & 1996 - 2020 & 1.82 & 2.54 & 4.36\\
Romania & 1996 - 2019 & 4.58 & 2.54 & 7.11\\
Latvia & 1996 - 2019 & 3.45 & 4.30 & 7.75\\
Turkey & 2009 - 2017 & -5.54 & 7.20 & 1.66\\
Russia & 2008 - 2018 & 1.36 & -1.44 & -0.08\\
\addlinespace
Uruguay & 2012 - 2016 & 4.93 & 1.36 & 6.29\\
Malaysia & 2006 - 2015 & 29.82 & 12.36 & 42.18\\
Kazakhstan & 1996 - 2019 & 2.71 & 6.79 & 9.50\\
Costa Rica & 2012 - 2017 & 7.98 & 5.21 & 13.18\\
Bulgaria & 1996 - 2017 & 4.01 & -0.09 & 3.91\\
\addlinespace
Chile & 1996 - 2009 & 14.69 & 6.72 & 21.41\\
Mexico & 1996 - 2019 & 8.94 & 3.28 & 12.23\\
Domin. Rep. & 1996 - 2016 & 29.14 & 8.44 & 37.58\\
Serbia & 1997 - 2019 & 16.60 & 9.62 & 26.22\\
Brazil & 1996 - 2018 & 7.58 & 4.44 & 12.02\\
\addlinespace
Colombia & 1996 - 2019 & 33.70 & 8.19 & 41.89\\
Azerbaijan & 1996 - 2010 & 38.33 & 13.76 & 52.09\\
Iran & 1996 - 2018 & -22.66 & 6.93 & -15.73\\
Peru & 2007 - 2019 & 7.87 & 6.44 & 14.30\\
Moldova & 1998 - 2019 & 11.97 & 4.15 & 16.11\\
\addlinespace
Egypt & 1996 - 2015 & 64.87 & 14.11 & 78.98\\
Mongolia & 2005 - 2019 & 2.19 & 6.86 & 9.05\\
Botswana & 1996 - 2000 & 15.19 & 6.70 & 21.89\\
Venezuela & 1997 - 2007 & 32.78 & 6.69 & 39.47\\
Guatemala & 2001 - 2019 & 22.59 & 6.98 & 29.57\\
\addlinespace
Cape Verde & 2007 - 2017 & 3.75 & 3.82 & 7.56\\
Uzbekistan & 2015 - 2017 & 2.64 & 6.99 & 9.63\\
Nicaragua & 2006 - 2018 & 5.67 & 5.11 & 10.78\\
Honduras & 2000 - 2015 & 16.45 & 8.54 & 24.99\\
Kyrgyzstan & 1996 - 2019 & 13.01 & 4.09 & 17.09\\
\addlinespace
Côte d’Ivoire & 1996 - 2000 & 50.56 & 29.54 & 80.10\\
Croatia & 1996 - 2019 & 92.44 & 4.24 & 96.67\\
Senegal & 2014 - 2015 & 23.29 & 8.68 & 31.97\\
Niger & 1996 - 2019 & 65.65 & 11.87 & 77.52\\
\bottomrule 
\end{tabular}
}
}
\label{tbl-fgr_table}
\end{table}

\FloatBarrier

\section{High values of \(f(K)\) and \(r(K)\) in some countries}

Table \ref{tbl-fgr_table} shows average values of \(f(K)\), calculated as \(\frac{C-\Pi}{K}\), as greater than 10\% in 26 of the 68 countries reported. 
Evidence from stock markets and business experience suggests that cash flow rates greater than 10\%, with concomitantly high rates of return, are unlikely at a national scale. Rather, we suspect that there are conditions under which national accounts tend to overstate consumption, or understate pay and market value of capital, or all at once. Our displays list countries in descending order of net income per capita due to a first impression, which may be wrong, that conditions for misgauging \(C, \Pi\) and \(K\) may have something to do with technological advancement\footnote{Only 5 of those 26 countries appear in the top half of all 68 countries ranked by net income per capita.}. Results for those 26 countries remind us that measurements in accounts, including national accounts, are generally imperfect, and that our derivation of \(f(K), g(K)\) and \(r(K)\) from measurements of \(C, \Pi\) and \(K\) cannot give exact results in any country at any time.

\section{The possibility of negative values of $\Delta K$ and $P$}

It is expected that \(\Delta K\), even at national or global scales, will occasionally be negative in times of recession, depression or external shocks. When \(- \Delta K\) exceeds \(F(K)\) absolutely, then net profit $P$ will also be negative. Table~\ref{tbl-fgr_table} and line charts in Section 1 of the \href{https://web-appendix.quarto.pub/separate-way-to-measure-rate-of-return/}{web appendix}\footnote{Web Appendix: https://web-appendix.quarto.pub/separate-way-to-measure-rate-of-return/} show these recurring realities of the investment world.

\section{Cash flow, consumption and pay}

Net profit might be spent either on investment or consumption. Cash flow, however, is defined as net of any concurrent investment, whether the source of that investment is from pay, or from transfer payments received, or reinvestment of cash flow received. Thus the only possible dispositions of cash flow at the household scale, where that set includes all existing or currently acquired assets of the household, are in consumption or in transfer payments to other households. If net transfer $\Gamma$ means transfer paid out less transfer received, then, this reasoning gives
\begin{equation}
\Pi + F(K) = C + \Gamma \quad \text{at the household scale,} \label{eq-13}
\end{equation}
as the source and disposition of payments. At the collective scale, where transfers offset to zero, then, 
\begin{equation}
\Pi + F(K) = C \ , \quad \text{from which} \quad C - \Pi = F(K) \ , \quad \text{collectively.} \label{eq-14}
\end{equation}
This confirms Eq. \eqref{eq-4}, and also Eqs. \eqref{eq-1} and \eqref{eq-2} by the reasoning shown in Eq. \eqref{eq-3}.

\section{Where our method fits}

We see our indirect derivation method for finding \(f(K)\) and \(r(K)\) as complementary to methods of finding them by direct research as in \cite{jorda2019}, which we choose as our current reference standard for that method. Table \ref{tbl-jorda_table} compares our findings with theirs. Their research approach gives finer detail, while ours by inference from \(C,\Pi\) and \(K\), subject to mismeasurement problems described in Section 3, gives broader coverage. Ours describes for capital as a whole, including owner-occupied housing and government property and personal effects, even if we are not sure what \(f(K)\) and \(r(K)\) mean for them. Meanwhile the research approach, which must focus on areas where assets are commonly rented or traded, illuminates those areas in ways which ours cannot, and with an accuracy that we cannot claim. Neither consumption, nor pay, nor market-value capital can be measured as closely as equity prices and dividend rates in finding \(f(K)\), \(g(K)\) and \(r(K)\) for the corporate sector.

\citeauthor{jorda2019} find \(f(K)\) separately for investments in equities, rental housing, and government bills and bonds as dividend income, rental income, and real interest rates respectively. Table \ref{tbl-jorda_table} shows these findings along with our derivations of \(f(K)\), \(g(K)\), and \(r(K)\) for each country for which \citeauthor{jorda2019} report. In those 16 countries, our impression is that a weighted average of the four measures of \(f(K)\) in \citeauthor{jorda2019} would come close to our single one derived from consumption, pay, and market-value capital, which in these countries seems to hold to a range of some 3\% to 6\% per year.

\begin{table}
\caption{Comparison of findings from derivation and market research(\%).}
\begin{tabular}{lccccccccccc}
\toprule
& \multicolumn{3}{c}{Derivation from \(C, \Pi\) and \(K\)} & \multicolumn{8}{c}{Market Research by \citeauthor{jorda2019}} \\
\cmidrule(lr){2-4} \cmidrule(lr){5-12} 
& \shortstack{\(\frac{C - \Pi}{K}\)} & \shortstack{\(\frac{\Delta K}{K}\)} &  \shortstack{\(\frac{C + \Delta K - \Pi}{K}\)} & \multicolumn{3}{c}{Equities} & \multicolumn{3}{c}{Housing} & \multicolumn{2}{c}{Gvt. Debt} \\
\cmidrule(lr){2-4} \cmidrule(lr){5-7} \cmidrule(lr){8-10} \cmidrule(lr){11-12}
Country & \(f(K)\) & \(g(K)\) & \(r(K)\) & \(\shortstack{Div\\Yld}\) & \(\shortstack{Cap\\Appr}\) & \(\shortstack{Total\\Rtn}\) & \(\shortstack{Rent\\Yld}\) & \(\shortstack{Cap\\Appr}\) & \(\shortstack{Total\\Rtn}\) & Bills & Bonds\\
\midrule
Australia & 2.40 & 4.45 & 6.85 & 4.90 & 3.06& 7.79& 3.99& 2.53& 6.37& 1.29& 2.24\\
Belgium & 1.54 & 2.80 & 4.33 & 3.83& 2.53& 6.23& 6.15& 1.95& 7.89& 1.21& 3.01\\
Denmark & 2.16 & 5.42 & 7.59 & 4.95& 2.71& 7.49& 7.13& 1.26& 8.22& 3.08& 3.58\\
Finland & 3.79 & 4.06 & 7.85 & 5.08& 5.19& 10.03& 7.14& 2.82& 9.58& 0.64& 3.22\\
France & 2.39 & 4.28 & 6.67 & 3.73& -0.37& 3.21& 5.09& 1.55& 6.39& -0.47& 1.54\\
\addlinespace
Germany & 3.53 & 3.11 & 6.64 & 4.08& 2.74& 7.11& 6.03& 1.86& 7.82& 1.51& 3.15\\
Italy & 4.89 & 2.84 & 7.73 & 3.61& 3.78& 7.25& 3.49& 1.45& 4.77& 1.20& 2.53\\
Japan & 2.28 & 2.81 & 5.09 & 2.65& 3.12& 6.00& 4.70& 2.00& 6.54& 0.68& 2.54\\
Netherlands & 2.42 & 3.35 & 5.77 & 4.87& 3.38& 8.10& 5.96& 1.75& 7.51& 1.37& 2.71\\
Norway & 3.55 & 5.64 & 9.19 & 4.21& 1.61& 5.67& 6.72& 1.49& 8.03& 1.10& 2.55\\
\addlinespace
Portugal & 2.77 & 2.54 & 5.32 & 2.28& 2.92& 5.11& 4.47& 1.13& 5.21& 0.01& 2.23\\
Spain & 2.72 & 3.57 & 6.28 & 4.53& 1.80& 5.83& 4.16& 1.26& 5.21& -0.04& 1.41\\
Sweden & 2.59 & 4.25 & 6.84 & 4.12& 4.08& 8.02& 7.12& 1.39& 8.30& 1.77& 3.25\\
Switzerland & 0.88 & 3.70 & 4.58 & 3.20& 3.17& 6.27& 4.54& 0.81& 5.24& 0.89& 2.41\\
UK & 5.11 & 3.00 & 8.11 & 4.53& 2.48& 6.83& 3.94& 1.63& 5.44& 1.16& 2.29\\
USA & 4.44 & 3.56 & 7.99 & 4.38& 4.19& 8.46& 5.33& 0.90& 6.10& 2.23& 2.85\\
\bottomrule
\multicolumn{12}{l}{\footnotesize Comments:}\\
\multicolumn{12}{l}{\footnotesize 1. Data for equities, housing, bills, and bonds are reproduced from \cite{jorda2019}, Tables IX and X.}\\
\multicolumn{12}{p{\dimexpr \textwidth-2\tabcolsep}}{\footnotesize 2. All data are shown from inception through 2015. Dates of inception are generally earlier for the data from \citeauthor{jorda2019} than from ours. Ours differ from those shown in our Table 1, which do not end in 2015. Exact periods of coverage of data from \cite{jorda2019} are shown in Table \ref{tbl-fgr_table} of that source.}
\end{tabular}
\label{tbl-jorda_table}
\end{table}

\section{Factor shares in consumption}

At the scale of all assets together, where $F(K)$ can no longer be reinvested in other capital, $F(K)$ simplifies to the cost of consumption afforded from net profit, here notated $C(K)$, as distinct from the part of consumption afforded from pay. Then 
\begin{equation}
F(K) = C(K) \quad \text{collectively.} \label{eq-7}
\end{equation}
It appears from Eq. \eqref{eq-4}, which can be arranged as $C = \Pi + F(K)$, that pay equals the remaining part of consumption afforded from pay rather than from cash flow $F(K)$. If this part is notated $C(\Pi)$, we have
\begin{equation}
\Pi = C(\Pi) \quad \text{and} \quad C = C(\Pi) + C(K) \quad \text{collectively,} \label{eq-8}
\end{equation}
to give the factor shares in consumption. Division by \(C\) gives
\begin{equation}
\frac{C(\Pi)}{C} + \frac{C(K)}{C} = \frac{\Pi}{C} + \frac{F(K)}{C} = 1 \, \quad \text{collectively,} \label{eq-9}
\end{equation}
to give those factor shares as percents. Results are shown in Table \ref{tbl-shares_table} below.

\FloatBarrier
\begin{table}[pos = H]
\caption{Labor and capital shares\(^a\) in consumption in 68 countries in descending order of income per capita(\%).}%
\makebox[\textwidth][c]{
{\centering 
\begin{tabular}{llll} \toprule
Country & Period & Labor Share & Capital Share\\
\midrule
Luxembourg & 1996 - 2018 & 102.27 & -2.27\\
Norway & 1981 - 2020 & 87.02 & 12.98\\
Netherlands & 1996 - 2019 & 96.89 & 3.11\\
Switzerland & 1995 - 2019 & 96.02 & 3.98\\
United States & 1970 - 2018 & 84.66 & 15.34\\
\addlinespace
Denmark & 1996 - 2020 & 91.98 & 8.02\\
Austria & 1996 - 2019 & 93.38 & 6.62\\
Sweden & 1950 - 2020 & 95.75 & 4.25\\
Ireland & 1996 - 2019 & 94.83 & 5.17\\
Saudi Arabia & 2002 - 2009 & 48.68 & 51.32\\
\addlinespace
Belgium & 1996 - 2019 & 91.36 & 8.64\\
Iceland & 2000 - 2014 & 89.31 & 10.69\\
Germany & 1990 - 2020 & 88.09 & 11.91\\
Cyprus & 1996 - 2019 & 77.00 & 23.00\\
France & 1950 - 2019 & 94.53 & 5.47\\
\addlinespace
Australia & 1960 - 2019 & 87.72 & 12.28\\
Canada & 1972 - 2020 & 91.46 & 8.54\\
Malta & 1996 - 2019 & 94.40 & 5.60\\
Finland & 1996 - 2020 & 87.92 & 12.08\\
Israel & 2000 - 2019 & 74.67 & 25.33\\
\addlinespace
UK & 1970 - 2019 & 78.76 & 21.24\\
New Zealand & 1999 - 2017 & 75.03 & 24.97\\
Aruba & 1996 - 2000 & 102.47 & -2.47\\
Spain & 1995 - 2019 & 86.21 & 13.79\\
Japan & 1980 - 2017 & 87.41 & 12.59\\
\addlinespace
Slovenia & 1996 - 2019 & 101.22 & -1.22\\
Lithuania & 1996 - 2019 & 74.04 & 25.96\\
Czechia & 1994 - 2019 & 92.74 & 7.26\\
Italy & 1980 - 2020 & 78.39 & 21.61\\
Estonia & 1996 - 2019 & 96.67 & 3.33\\
\addlinespace
Poland & 1996 - 2019 & 90.63 & 9.37\\
Slovakia & 1996 - 2020 & 82.40 & 17.60\\
Hungary & 1996 - 2019 & 103.77 & -3.77\\
Greece & 1996 - 2019 & 71.56 & 28.44\\
\bottomrule \end{tabular}
    }
    \centering{
        \begin{tabular}{llll} \toprule
Country & Period & Labor Share & Capital Share\\
\midrule
Portugal & 1996 - 2020 & 84.86 & 15.14\\
Romania & 1996 - 2019 & 78.04 & 21.96\\
Latvia & 1996 - 2019 & 76.76 & 23.24\\
Turkey & 2009 - 2017 & 107.30 & -7.30\\
Russia & 2008 - 2018 & 92.37 & 7.63\\
\addlinespace
Uruguay & 2012 - 2016 & 72.58 & 27.42\\
Malaysia & 2006 - 2015 & 65.28 & 34.72\\
Kazakhstan & 1996 - 2019 & 87.29 & 12.71\\
Costa Rica & 2012 - 2017 & 75.03 & 24.97\\
Bulgaria & 1996 - 2017 & 74.70 & 25.30\\
\addlinespace
Chile & 1996 - 2009 & 71.80 & 28.20\\
Mexico & 1996 - 2019 & 62.36 & 37.64\\
Domin. Rep. & 1996 - 2016 & 68.22 & 31.78\\
Serbia & 1997 - 2019 & 84.14 & 15.86\\
Brazil & 1996 - 2018 & 75.55 & 24.45\\
\addlinespace
Colombia & 1996 - 2019 & 69.33 & 30.67\\
Azerbaijan & 1996 - 2010 & 64.32 & 35.68\\
Iran & 1996 - 2018 & 126.22 & -26.22\\
Peru & 2007 - 2019 & 71.48 & 28.52\\
Moldova & 1998 - 2019 & 81.18 & 18.82\\
\addlinespace
Egypt & 1996 - 2015 & 63.44 & 36.56\\
Mongolia & 2005 - 2019 & 87.01 & 12.99\\
Botswana & 1996 - 2000 & 63.68 & 36.32\\
Venezuela & 1997 - 2007 & 70.68 & 29.32\\
Guatemala & 2001 - 2019 & 56.99 & 43.01\\
\addlinespace
Cape Verde & 2007 - 2017 & 96.49 & 3.51\\
Uzbekistan & 2015 - 2017 & 85.24 & 14.76\\
Nicaragua & 2006 - 2018 & 71.07 & 28.93\\
Honduras & 2000 - 2015 & 74.30 & 25.70\\
Kyrgyzstan & 1996 - 2019 & 78.03 & 21.97\\
\addlinespace
Côte d’Ivoire & 1996 - 2000 & 80.85 & 19.15\\
Croatia & 1996 - 2019 & 13.10 & 86.90\\
Senegal & 2014 - 2015 & 74.34 & 25.66\\
Niger & 1996 - 2019 & 39.88 & 60.12\\
\bottomrule \end{tabular}
    }
}
\raggedright{\footnotesize \(^a\) Derived at \(\frac{\Pi}{C}\) and \(\frac{C-\Pi}{C}\) respectively.}
\label{tbl-shares_table}
\end{table}

\FloatBarrier

Section 2 of the \href{https://web-appendix.quarto.pub/separate-way-to-measure-rate-of-return/}{web appendix}\footnote{See footnote 7 for web address.} to this paper shows line charts plotting labor and capital shares in consumption for each country over time. Note that only 10 of the 34 countries ranked in the top half of these 68 countries by net income per capita, but 32 of the 34 ranked in the lower half, show labor shares in consumption less than 80\%. We suspect that relative undermeasurement of pay, as suggested in explaining high findings for \(f(K)\) in less advanced economies in Table 1, may account for at least part of this difference (See Appendix A.4.).

\section{Rate of return in the stationary state}
The ratio \(C(K)/K\) can be shown as \(c(K)\). Then Eq. \eqref{eq-7} can be divided by \(K\) to show
\begin{equation}
f(K) = c(K) \quad \text{collectively.}\label{eq-10}
\end{equation}
Substitution in Eq. \eqref{eq-6} now gives
\begin{equation}
r(K) = g(K) + c(K) \quad \text{collectively.}\label{eq-11}
\end{equation}
The classical concept of the stationary state means a state where \(g(K)\) holds at zero. Eq. \eqref{eq-11} shows that collective return in the stationary state holds at \(c(K)\), the ratio of consumption drawn from capital to capital. Thus 
\begin{equation}
r(K) = c(K) \quad \text{collectively, in the stationary state.}\label{eq-12}
\end{equation}
A qualifier is appropriate here. By rate of return, we and \citeauthor{jorda2019} mean unleveraged return on capital owned free and clear of all debt. Some economists have meant return in an opposite sense as return to an imaginary entrepreneur who has invested no equity, but rather has borrowed the whole cost of capital employed. Rate of return to such an entrepreneur would tend to equal zero in the stationary state. Eq. \eqref{eq-12} applies only in the deleveraged sense used by us, by \citeauthor{jorda2019}, and by most investors.

\section{Data Sources}

The data for this study other than those reporting market research are sourced from the World Inequality Database (WID), accessible online at \href{https://wid.world}{WID.world}. This comprehensive database integrates data from national accounts and taxation records across 105 countries, standardizing these to align with the United Nations System of National Accounts (SNA) guidelines. For our analysis, we specifically utilize data from the subset of countries that provide comprehensive reports on net income, labor income share, consumption and market-value capital.

The primary variables utilized from WID include net national income (\texttt{mnninc}), the share of labor income (\texttt{wlabsh}), which is represented by us as \(\Pi\), and purchasing power parity conversions of local currencies to USD (\texttt{xlcusp}). WID estimates labor share as compensation of employees plus 70\% of mixed income. Additionally, we use average national income per capita (\texttt{anninc}) for arranging countries in our tables and graphs, and market-value national wealth (\texttt{mnweal}) as a representation of wealth $K$. Consumption $C$, as referenced in our analyses, aggregates Government Final Consumption Expenditure (\texttt{mcongo}) and Private Expenditures of Households and NPISH (\texttt{mconhn}), encapsulating both government expenditure (GCE) and personal consumption expenditure (PCE).

Data reporting market research are directly sourced from the tables presented in \cite{jorda2019}.

\section{Discussion and conclusions}

Our method of deriving cash flow rate, capital growth rate, and rate of return from data in national accounts serves as a complement to the traditional method relying on direct research, as in the work of \cite{jorda2019}. Our results tallied broadly with theirs in the 16 countries for which they report.

The labor and capital shares shown in Table \ref{tbl-jorda_table} are broadly consistent with expectations. The labor share of net product is usually estimated at 70\% or so. As consumption is smaller than net product whenever capital growth is positive, the labor share of consumption applies the same numerator (pay) to a smaller denominator (consumption versus net product). Also see Appendix A.4..

\customlabel{appendix-a}{A}{%
\section*{Appendix A. \hspace{0.5em} Output in terms of human capital}}

\renewcommand{\theequation}{A.\arabic{equation}}
\setcounter{equation}{0}

\customlabel{appendix-a1}{A.1}{%
\subsection*{Appendix A.1. \hspace{0.5em} Necessary adjustments to the doctrine \(Y=\Delta K + C\)}}
We take value $V$ to mean the sum of human capital $H$ and physical capital $K$. Then
\begin{equation}
V = H + K\ . \label{eq-a1}
\end{equation}
Value growth $\Delta V$ equals value creation $Y$, also called net output or value added, plus investment from outside, less value yielded out. Value yielded out less value invested in from outside is cash flow $F(V)$, which may include cash flow from human capital (see Eq. \ref{eq-a3} below). Then
\begin{equation}
\Delta V = Y - F(V) \ , \quad \text{from which} \quad Y = \Delta V + F(V) \ . \label{eq-a2}
\end{equation}

The sum $\Delta V + F(V)$, or value growth plus cash flow, is called total return. Thus Eq. \eqref{eq-a2} shows the equivalence of value added and total return.
Although human capital is usually measured at cumulative cost, we follow \cite{petty1664verbum}, Fisher (\citeyear{fisher1907}, \citeyear{fisher1930}), \cite{mincer1958} and others in measuring it at present value, or expected lifetime cash flow discounted at the time preference rate. The flow discounted is pay, including imputed pay, as the flow yielded out by human capital to its human owner, less invested consumption\footnote{The concepts of invested consumption, pure consumption, self-invested work and human depreciation were introduced in \cite{schultz1961}.} as the flow invested in human capital from outside itself. If invested consumption shows as $C_s$, then, we find
\begin{equation}
F(H) = \Pi - C_s \ , \label{eq-a3}
\end{equation}
where $F(H)$ is cash flow of human capital. We also follow \cite{ben-porath1967} in describing the growth of human capital as the sum of invested consumption plus self-invested work less human depreciation. If the last two flows show as $W_s$ and $D(H)$, we have\footnote{Our terms and notations need not follow those of our sources. We are not aware, for example, that others have used the term “human cash flow”.}
\begin{equation}
\Delta H = C_s + W_s - D(H) \ . \label{eq-a4}
\end{equation}
Eq. \eqref{eq-a2} can be applied to human capital separately to find work $W$, the net output produced by human capital, as 
\begin{align}
W = \Delta H + F(H) = C_s + W_s - D(H) + \Pi - C_s = \Pi + W_s - D(H) \ , \label{eq-a5}
\end{align}
showing that work exceeds pay by the amount $W_s - D(H)$.

Cash flow from assets generally can be spent on investment in other assets or on consumption. Cash flow from all physical capital of all households together is net of all investment in physical capital, and is spent wholly on consumption (Eq. \eqref{eq-7}). Cash flow from physical and human capital together is net of all investment in human capital through invested consumption, and so is spent on pure consumption alone\footnote{The flow which maintains rather than creates human capital, and so is exhausted from the economy in satisfying tastes for vitality and survival.}. If this is notated $C_p$, then preceding equations allow
\begin{equation}
\begin{aligned}
Y &= \Delta V + C_p \\
&= \Delta H + \Delta K + C_p \\
&= C_s + W_s - D(H) + \Delta K + C_p \quad \text{collectively.} \label{eq-a6}
\end{aligned}
\end{equation}
As consumption $C$ is either invested in human capital or yielded from the economy in satisfying tastes, we may write
\begin{equation}
C_s + C_p = C \ . \quad \text{From Eq. \eqref{eq-a6}, then} \label{eq-a7}
\end{equation}
\vspace{-5ex}
\begin{equation}
Y = C + W_s - D(H) + \Delta K \ , \quad \text{collectively,} \label{eq-a8}
\end{equation}
showing that $Y$ exceeds the sum $C + \Delta K$ by the amount $W_s - D(H)$ at the collective scale.

\customlabel{appendix-a2}{A.2}{%
\subsection*{Appendix A.2. \hspace{0.5em} Confirmation of Eq. (\ref{eq-1})}}

Net output can also be found at the sum of net outputs of the factors. That is,
\begin{equation}
Y = W + P \ . \quad \text{Substitution of Eq. \eqref{eq-a5} in this gives} \label{eq-a9}
\end{equation}
\vspace{-5ex}
\begin{equation}
Y = \Pi + W_s - D(H) + P \ . \quad \text{This and Eq. \eqref{eq-a8}, with rearrangement, allow} \label{eq-a10}
\end{equation}
\vspace{-5ex}
\begin{equation}
Y = C + \Delta K + W_s - D(H) = \Pi + P + W_s - D(H) \ , \quad \text{collectively,} \label{eq-a11}
\end{equation}
from which
\begin{equation}
C + \Delta K = \Pi + P \ , \quad \text{collectively,} 
\end{equation}
confirming Eq. \eqref{eq-1} at the collective scale.

\customlabel{appendix-a3}{A.3}{%
\subsection*{Appendix A.3. \hspace{0.5em} Factor shares in net income}}

We have treated the larger factor as human capital measured as present value of prospective lifetime pay less invested consumption, rather than as labor measured, say, in headcount or hours or current pay, but we follow custom by terming its share in net income or in consumption as the labor share. Eq. (A.11) shows that net output, and equivalently net income, exceed the sums $C + \Delta K$, or equivalently $\Pi + P$, by the difference $W_s - D(H)$.

The labor share in consumption was reasoned in Eqs. \eqref{eq-8} and \eqref{eq-9} to equal $\Pi$ absolutely, and $\Pi / C$ as a percentage. The capital share in consumption was reasoned to equal $F(K)$ absolutely, and $F(K)/C$ as a percentage, since this is the only remaining possible disposition of $C$. Thus measurement of the factor shares in $C$ would require only measurement of $\Pi$ and $C$, where measurement of $F(K)$ is inferred as their difference rather than measured separately.

Factor shares in net income, however, would be defined as $W$ and $P$ absolutely and as $W/Y$ and $P/Y$ as percentages. Equations above allow
\begin{equation}
\frac{W}{Y} = \frac{\Pi + W_s - D(H)}{C + \Delta K + W_s - D(H)} \quad  \text{and} \quad \frac{P}{Y} = \frac{\Delta K + C - \Pi}{C + \Delta K + W_s - D(H)} \ , \label{eq-a13}
\end{equation}
showing that measurement of factor shares in net income require measurement of $K,\ W_s$ and $D(H)$ as well as $C$ and $\Pi$. As human capital leaves little market record apart from its rise in pay, and from an important component of invested consumption in the cost of schooling, measurements of $Y$ and $W$ are problematic. We suggest that distribution theory might consider a refocus from shares in net income to shares in consumption on these grounds of measurability and others. It can be argued that individuals vary in appetite for wealth, and in talent for managing wealth, and in willingness to accept the responsibility for that management, but that all agree in need for consumption. Benevolence and wisdom of public  policy, if so, might better be measured by ample sufficiency of consumption than by equality of net income, so that individual tastes and talents can account for the rest of differences in net income.

\customlabel{appendix-a4}{A.4}{%
\subsection*{Appendix A.4. \hspace{0.5em} Summary and discussion}}

The bad news is that we cannot measure net output in the sense of value added to both factors. The sums we had thought measured it, meaning $C + \Delta K$ or equivalently $\Pi + P$, both overlook $W_s$ and $D(H)$ as positive and negative components in value added to human capital. Neither $W_s$ (self-invested work) nor $D(H)$ (human depreciation) is practical to measure.

The good news is that we can measure net profit at market value, including imputed net profit on the housing and government sectors, subject to measurement problems discussed earlier, from data for consumption, pay and market value capital alone. Meanwhile we can measure the labor and capital shares in consumption as $\Pi / C$ and $( C - \Pi ) / C$, with the same qualifier, without need for other data. This convenience gives some compensation for the difficulty in finding net income, or factor shares in net income as adjusted for $W_s$ and $D(H)$.

As arguments determining \(f(K), g(K), r(K), c(\Pi)\) and \(c(K)\) relied on no simplifying assumptions, we could find those values exactly if we could measure \(C, \Pi\) and \(K\) exactly. We cannot. It may be possible, however, to focus more attention on measuring them, and in seeking consensus as to appropriate imputations where measurement is impractical. It may be the case, for example, that pay tends to be less fully recognized in agrarian economies, where higher percentages of people are self-employed homemakers, farmers and herders. Capital too might tend to be relatively underreported if such people are more likely to make implements for use by themselves and their communities, and to construct their own housing, rather than buy and sell those things in formal markets. If any such differences in completeness of measurement were verified, and quantified in standard imputations, it might be possible to reach more realistic versions of Tables \ref{tbl-fgr_table} and \ref{tbl-jorda_table} for those countries and others.

\printcredits

\bibliographystyle{cas-model2-names}

\bibliography{cas-refs}

\begin{thebibliography}{8}
\expandafter\ifx\csname natexlab\endcsname\relax\def\natexlab#1{#1}\fi
\providecommand{\url}[1]{\texttt{#1}}
\providecommand{\href}[2]{#2}
\providecommand{\path}[1]{#1}
\providecommand{\DOIprefix}{doi:}
\providecommand{\ArXivprefix}{arXiv:}
\providecommand{\URLprefix}{URL: }
\providecommand{\Pubmedprefix}{pmid:}
\providecommand{\doi}[1]{\href{http://dx.doi.org/#1}{\path{#1}}}
\providecommand{\Pubmed}[1]{\href{pmid:#1}{\path{#1}}}
\providecommand{\bibinfo}[2]{#2}
\ifx\xfnm\relax \def\xfnm[#1]{\unskip,\space#1}\fi
\bibitem[{Ben-Porath(1967)}]{ben-porath1967}
\bibinfo{author}{Ben-Porath, Y.}, \bibinfo{year}{1967}.
\newblock \bibinfo{title}{The production of human capital and the life cycle of
  earnings}.
\newblock \bibinfo{journal}{Journal of Political Economy} \bibinfo{volume}{75},
  \bibinfo{pages}{352--365}.
\bibitem[{Fisher(1907)}]{fisher1907}
\bibinfo{author}{Fisher, I.}, \bibinfo{year}{1907}.
\newblock \bibinfo{title}{The Rate of Interest}.
\newblock \bibinfo{publisher}{McMaster University Archive for the History of
  Economic Thought}.
\bibitem[{Fisher(1930)}]{fisher1930}
\bibinfo{author}{Fisher, I.}, \bibinfo{year}{1930}.
\newblock \bibinfo{title}{The Theory of Interest, as Determined by Impatience
  to Spend Income and Opportunity to Invest It}.
\newblock \bibinfo{publisher}{Macmillan}, \bibinfo{address}{New York}.
\bibitem[{Homer and Sylla(2005)}]{homer2005history}
\bibinfo{author}{Homer, S.}, \bibinfo{author}{Sylla, R.}, \bibinfo{year}{2005}.
\newblock \bibinfo{title}{A History of Interst Rates}.
\newblock Wiley finance, \bibinfo{publisher}{Wiley}.
\newblock \bibinfo{note}{Tex.lccn: 2005281578}.
\bibitem[{Jordà et~al.(2019)Jordà, Knoll, Kuvshinov, Schularick and
  Taylor}]{jorda2019}
\bibinfo{author}{Jordà, O.}, \bibinfo{author}{Knoll, K.},
  \bibinfo{author}{Kuvshinov, D.}, \bibinfo{author}{Schularick, M.},
  \bibinfo{author}{Taylor, A.M.}, \bibinfo{year}{2019}.
\newblock \bibinfo{title}{{The rate of return on everything, 1870–2015}}.
\newblock \bibinfo{journal}{The Quarterly Journal of Economics}
  \bibinfo{volume}{134}, \bibinfo{pages}{1225--1298}.
\newblock \DOIprefix\doi{10.1093/qje/qjz012}.
\bibitem[{Mincer(1958)}]{mincer1958}
\bibinfo{author}{Mincer, J.}, \bibinfo{year}{1958}.
\newblock \bibinfo{title}{Investment in human capital and personal income
  distribution}.
\newblock \bibinfo{journal}{Journal of Political Economy} \bibinfo{volume}{66},
  \bibinfo{pages}{281--302}.
\bibitem[{Petty(1664)}]{petty1664verbum}
\bibinfo{author}{Petty, W.}, \bibinfo{year}{1664}.
\newblock \bibinfo{title}{Verbum Sapienti}.
\bibitem[{Schultz(1961)}]{schultz1961}
\bibinfo{author}{Schultz, T.W.}, \bibinfo{year}{1961}.
\newblock \bibinfo{title}{Investment in human capital}.
\newblock \bibinfo{journal}{The American Economic Review} \bibinfo{volume}{51},
  \bibinfo{pages}{1--17}.

\end{thebibliography}

\end{document}